%% file: main.tex
\documentclass{fizik}
\usepackage{hyperref}
\hypersetup{
colorlinks=true,
urlcolor=blue,
citecolor=blue}
\usepackage[all]{xy,xypic}
\usepackage{amsfonts,amssymb,amsmath,amsgen,amsopn,amsbsy,theorem,graphicx,epsfig}
\usepackage{eufrak,amscd,bezier,latexsym,mathrsfs,enumerate,multirow}
\usepackage[utf8]{inputenc}\usepackage[english]{babel}
\usepackage[dvipsnames]{xcolor}
\usepackage{academicons}
\usepackage{tabularx}
\usepackage{caption}
\definecolor{orcidlogocol}{HTML}{A6CE39}

\usepackage[pagewise]{lineno}

\usepackage{changes}

\yil{2021}
\vol{45}
\fpage{247}
\lpage{267}
\doi{10.3906/fiz-2110-16}
\title{Deriving Weak Deflection Angle by Black Holes or Wormholes using Gauss-Bonnet Theorem }

\author[ YASHMITHA KUMARAN \& AL\.{I} \"{O}VG\"{U}N]{
\textbf{YASHMITHA KUMARAN$^{1}$ \& AL\.{I} \"{O}VG\"{U}N$^{1}$  \thanks{ali.ovgun@emu.edu.tr}~\href{https://orcid.org/0000-0002-9889-342X}{}}\\
\\ 
$^{1}$Physics Department, Eastern Mediterranean
University, Famagusta, North Cyprus \\ 99628 via Mersin 10, Turkey
\\ [1.8em]

\rec{20.10.2021}
\acc{22.10.2021}
\finv{02.11.2021}
}

\input{fizik.tex}

\begin{document}

\maketitle

\begin{abstract}In this review, various researches on finding the bending angle of light deflected by a massive gravitating object which regard the Gauss-Bonnet theorem as the premise have been revised. Primarily, the Gibbons and Werner method is studied apropos of the gravitational lensing phenomenon in the weak field limits. Some exclusive instances are deliberated while calculating the deflection angle, beginning with the finite-distance corrections on non-asymptotically flat spacetimes. Effects of plasma medium is then inspected to observe its contribution to the deflection angle. Finally, the Jacobi metric is explored as an alternative method, only to arrive at similar results. All of the cases are probed in three constructs, one as a generic statement of explanation, one for black holes, and one for wormholes, so as to gain a perspective on every kind of influence.

\end{abstract}

\section{Introduction}
Light is a unique form of radiation: from vision to life, it serves its purpose in various ways. The most interesting aspect of light, however, is the way it interacts with heavy objects. Added to solving the mystery behind the precession of Mercury, this bending of light in the presence of a massive object opened up a wide spectrum of prospects to explore, such as the spacetime and its null structure \cite{1}. In this paper, we will be focusing on a phenomenon called gravitational lensing that is a fascinating consequence of deflection of light.

Gravitational lensing occurs when an astronomical object is massive enough to bend the elapsing light into a lens, enabling the observer to capture more information about the source than feasible. Depending on the nature of the lens, it is classified as strong lensing or weak lensing. Here, we are concerned about weak lensing which is caused by marginal distortions since it singles out mass distributions due to minute magnifications.  Some entities of curiosity are dark energy, dark matter, exoplanets, galaxy clusters, black holes, wormholes, gravitational monopoles, etc. \cite{l1}-\cite{l56}. This established the foundations of general relativity which serve as the basis theoretical advancements in science till this day.

Walia et al. have studied the gravitational deflection of light due to rotating black holes in Horndeski gravity using Gauss-bonnet theorem in \cite{l1}. Qiao and Zhou have analyzed the gravitational deflection angle of light, weak gravitational lensing and Einstein ring using the Gauss-Bonnet theorem for an acoustic Schwarzschild black hole in \cite{l2}.
In \cite{l3}, the authors have discussed the effects of nonlinear electrodynamics on non-rotating black holes parametrized by the field coupling parameter and magnetic charge parameter using the Gauss-Bonnet theorem under the influence of the Generalized Uncertainty Principle.
Li and Jia have investigated the deflection of a charged particle moving in the equatorial plane of Kerr-Newman spacetime in the weak field limit in \cite{l4} using the Jacobi geometry and the Gauss-Bonnet theorem.
The authors have furthered their work with Liu to find the deflection and gravitational lensing of light and massive particles using the Gauss-Bonnet theoremin an arbitrary static, spherically symmetric and asymptotically (anti-)de Sitter spacetimes in \cite{l5}.
Javed et al. have calculated the deflection angle for a Kazakov-Solodukhin black hole with the Gauss-Bonnet theorem to comprehend the effects of the deformation parameter both in vacuum and in the presence of plasma \cite{l6}.
Carvalho et al. have considered the Casimir energy corrected by the Generalized Uncertainty Principle as the source in their work \cite{l7} to investigate the gravitational bending angle due to the Casimir wormholes with the Jacobi metric in the Gauss-Bonnet theorem.
Ali and Kaushal have modified the Kerr-Newman black holes to find the exact solutions of rotating black holes in Eddington-inspired Born-Infeld gravity \cite{l8}.
In \cite{l9}, Pantig et al. have discussed the effects of dark matter on a Schwarzschild black hole by means of the Extended Uncertainty Principle.
Takizawa and Asada have examined the methods for iterative solutions of the gravitational lens equations in the strong deflection limit in \cite{l10} with the case study of Sagittarius A* and M87.
The authors of \cite{l11} have studied the influence of the Lorentz symmetry-breaking in the bending angle of massive particles and light for bumblebee black hole solutions. Based on two types of Non-linear electrodynamic (NLED) models, Fu et al. have deliberated on two black hole solutions with the Euler-Heisenberg NLED model and the Bronnikov NLED model, and have calculated their weak deflection angles with the help of the Gauss-Bonnet theorem for vacuum and plasma in \cite{l12}.
The deflection of a massive, charged particle by a novel four-dimensional charged Einstein-Gauss-Bonnet black hole is examined by Li et al. in \cite{l13} based on the Jacobi metric method.
Gullu and \"Ovg\"un have tested the effect of the global monopole and the bumblebee fields causing the spontaneous Lorentz symmetry-breaking \cite{l14}.
The authors of \cite{l15} have investigated the gravitational lensing by asymptotically flat black holes in the framework of Horndeski theory in weak field limits using the Gauss-Bonnet theorem to find the deflection angle in vacuum and plasma medium.

A static spherically symmetric wormhole solution due to the vacuum expectation value of a Kalb-Ramond field is obtained by Lessa at al. in \cite{l16}.
Javed at al. have analyzed the weak gravitational lensing of the Einstein-non-linear-Maxwell-Yukawa black hole in their work \cite{l17} in vacuum and in the presence of plasma.
In \cite{l18}, the authors have shown that every light-like geodesic in the NUT (Newman, Unti and Tamburino) metric projects to a geodesic of a two-dimensional Riemannian metric a.k.a. the optical metric with the help of Fermat’s principle.
Moumni et al. have studied the gravitational lensing by some black hole classes within the non-linear electrodynamics in \cite{l19} for the weak field limits using the Gauss-Bonnet theorem for vacuum and plasma medium.
Weak gravitational lensing by Bocharova-Bronnikov-Melnikov-Bekenstein black hole is demonstrated by Javed et al. in \cite{l20} also for vacuum and plasma medium.
Arakida has proposed a novel concept of the total deflection angle of a light ray in terms of the optical geometry, i.e., the Riemannian geometry experienced by the light ray \cite{l21}.
Khan and Ren have explored the effects of quintessential dark energy and its consequences on the spacetime geometry of a black hole on horizons and the silhouette generated by a Kerr-Newman black hole in \cite{l22}.
In \cite{l23}, Takizawa et al. have discussed a gravitational lens for an observer and source located within a finite distance from a lens object without assuming asymptotic flatness.
Light bending caused by a slowly rotating source in quadratic theories of gravity with the Einstein–Hilbert action extended by additional terms quadratic in the curvature tensors is studied in \cite{l24} by Buoninfante and Giacchini.
\"Ovg\"un et al. have used a new asymptotically flat and spherically symmetric solution in the generalized Einstein-Cartan-Kibble-Sciama theory of gravity to determine the weak deflection angle using the Gauss-Bonnet theorem in \cite{l25}.
Islam et al. have generalized the work on non-trivial 4D Einstein-Gauss-Bonnet theory of gravity in their paper \cite{l26} for the gravitational lensing by a Schwarzschild black hole.
The authors of \cite{l27} have examined the weak gravitational lensing by stringy black holes for plasma and non-plasma mediums.
The deflection angle of a dirty black hole is presented in \cite{l28} by Pantig and Rodulfo; essentially, a Schwarzschild black hole of mass surrounded by the dark matter. An exact solution of Kerr black hole surrounded by a cloud of strings in Rastall gravity is obtained through the Newman-Janis algorithm in \cite{l31} by Li and Zhou for various cases by employing the Gauss-Bonnet theorem.

Li and \"Ovg\"un have studied the weak gravitational deflection angle of relativistic massive particles by the Kerr-like black hole in the bumblebee gravity model in the weak field limits in \cite{l32}.
The work of Li and Jia in \cite{l33} scrutinizes the weak gravitational deflection of relativistic massive particles for a receiver and source at finite distance from the lens in stationary, axisymmetric and asymptotically flat spacetimes by extending the generalized optical metric method to the generalized Jacobi metric method using the Jacobi-Maupertuis Randers-Finsler metric.
Crisnejo et al. have shown that the Gauss-Bonnet theorem can be applied to describe the deflection angle of light rays in plasma media in stationary spacetimes in \cite{l34} and also have obtained the leading order behavior of the deflection angle of massive/massless particles in the weak field regime with higher order corrections in a cold non magnetized plasma.
The equivalence of the Gibbons-Werner method to the standard geodesic method with the case study of Kerr-Newman spacetime especially for the asymptotically flat case in \cite{l35} by Li and Zhou.
\"Ovg\"un et al. have examined the light rays in a static and spherically symmetric gravitational field of null aether theory using the Gauss-Bonnet theorem to determine the deflection angle and showing that the bending of light stems from a global and topological effect \cite{l36}.
Weak gravitational lensing in the background of Kerr-Newman black hole with quintessential dark energy has been explored by Javed et al. and they have extended their work by finding the deflection angle of light for rotating charged black hole with quintessence in \cite{l37}.
They have also taken an interest in a model of exact asymptotically flat charged hairy black holes in the background of dilaton potential in \cite{l38} and have shown the effect of the hair on the deflection angle in weak field limits for vacuum and plasma medium.
The authors of \cite{l39} have analyzed the weak gravitational lensing in a plasma medium and computed the deflection angle of non-geodesic trajectories followed by relativistic test massive charged particles in a Reissner-Nordstr\"om spacetime as an application.
Leon and Vega have studied the weak-field deflection of light by different mass distributions \cite{l40}.
Roesch and Werner have applied the results of General Relativity on the isoperimetric problem to show that length-minimizing curves subject to an area constraint are circles, and have also discussed the implications for the photon spheres of Schwarzschild, Reissner-Nordstr\"om, as well as continuous mass models solving the Tolman-Oppenheimer-Volkoff equation in \cite{l41}.
In \cite{l42}, Crisnejo et al. have investigated the finite distance corrections to the light deflection in a gravitational field with a plasma medium.
A rotating global monopole is discussed by Ono et al. in \cite{l43} as a possible extension to an asymptotically non-flat spacetime of a method improved with a generalized optical metric to find the deflection of light for an observer and source at finite distance from a lens object in a stationary asymptotically flat spacetime. 
They have also calculated the gravitomagnetic bending angle of light using this in \cite{l44}.
The same method has been used by them in the weak field approximation to calculate the deflection angle for rotating Teo wormhole \cite{l45}.
\"Ovg\"un has also used this method in \cite{l46} applying it to the non-rotating and rotating Damour-Solodukhin wormholes spacetimes to explore the gravitational lensing effects of these objects.

In \cite{l47}, \"Ovg\"un et al. have found a new traversable wormhole solution in the framework of a bumblebee gravity model in which the Lorentz symmetry violation arises from the dynamics of a bumblebee vector field that is non-minimally coupled with gravity.
The authors have studied the weak deflection angle in the spacetime of rotating regular black hole \cite{l48}.
Weak gravitational lensing by black holes and wormholes in the context of massive gravity theory is scrutinized by the authors of \cite{l49} and they have established the time delay problem in the spacetime of black holes and wormholes, respectively.
Jusufi and \"Ovg\"un reported the effect of the cosmological constant and the internal energy density of a cosmic string on the deflection angle of light in the spacetime of a rotating cosmic string with internal structure \cite{l50}.
In \cite{l51}, Jusufi et al. have studied quantum effects on the deflection angle using the Gauss-Bonnet theorem in the spacetime of  global monopole and a cosmic string.
Arakida has re-examined the light deflection in the Schwarzschild and the Schwarzschild–de Sitter spacetimes in \cite{l52} so as to propose the definition of the total deflection angle of the light ray by constructing a quadrilateral on the optical reference geometry determined by the optical metric in a supposedly static and spherically symmetric spacetime.
An electrically charged traversable wormhole solution is given by Goulart for the Einstein-Maxwell-dilaton theory when the dilaton is a phantom field \cite{l53}.
The work of Jusufi and \"Ovg\"un in \cite{l54} shows the calculation of the quantum correction effects on the deflection of light in the spacetime geometry of a quantum improved Kerr black hole pierced by an infinitely long cosmic string.
Authors have investigated the Lorentz symmetry breaking effects on the deflection of light by a rotating cosmic string spacetime in the weak limit approximation in \cite{l55}.
The authors of \cite{l56} have probed the deflection of light by a rotating global monopole spacetime and a rotating Letelier spacetime in the weak deflection approximation.
Bloomer \cite{l57} has pursued a geometrical approach to gravitational lensing theory and has extended it to the axially symmetric Kerr spacetime arriving at an expression for the gravitational deflection angle in the equatorial plane.
Gibbons and Warnick have made use of the fact that the optical geometry near a static non-degenerate Killing horizon is asymptotically hyperbolic in their work \cite{l58} to investigate universal features of black hole physics.
In \cite{l59}, Gibbons et al. have considered a triality between the Zermelo navigation problem, the geodesic flow on a Finslerian geometry of Randers type, and spacetimes in one dimension higher admitting a time-like conformal Killing vector field.

In section \ref{sgbt}, a brief review of the Gauss-Bonnet Theorem (GBT) is given and in section \ref{sgwm}, the GW method, an extension of GBT proposed by Gibbons and Werner in 2008, is studied; these are the basis of finding the deflection angle of light in this paper. In section \ref{sfdm}, we have inspected the method to induce finite-distance corrections for non-asymptotically flat spacetimes while determining the deflection angle. Section \ref{spm} concentrates on the effects of homogenous plasma on the deflection angle as a medium instead of the light rays travelling through vacuum. Section \ref{ssbh} draws attention to stationary black holes, followed by section \ref{sjma} specifying the Jacobi Matrix approach, and the conclusion in section \ref{conc}.

\section{Brief Review of Gauss-Bonnet Theorem}
\label{sgbt}
The surface that the bending of light occurs in is the key to compute the weak deflection angle; the rays of light are treated as space-like geodesics of the optical metric. Introducing the Gauss-Bonnet theorem, an approach that utilizes these attributes by relating the topology of the surface to its intrinsic geometry, it facilitates the bending angle to be invariant under coordinate transformations according to \cite{4}.

To elaborate on this, let's look at the simplest example of a triangle as in \cite{5}. Say, the interior angles of the triangle are $\theta_a$, $\theta_b$ and $\theta_c$ so that: $\theta_a + \theta_b + \theta_c = \pi $.
Then, the corresponding exterior angles will be: $ \pi-\theta_a$, $\pi-\theta_b$, and $\pi-\theta_c$. \newline 
For a spherical triangle belonging to a unit sphere, the above equation changes to:
$ \theta_a + \theta_b + \theta_c > \pi $
such that the additional quantity is related to the area of the triangle as:
$ \theta_a + \theta_b + \theta_c - \pi = A_\Delta $.
Therefore, for a sphere of radius, $R$, this can be written as:$ \theta_a + \theta_b + \theta_c - \pi = \frac{A_\Delta}{R^2} $
\newline Applying the Gauss-Bonnet theorem over the given area with the Gaussian curvature of the sphere, $\mathcal{K}$:
\begin{align}
    \int_{A_\Delta} \mathcal{K} \mathrm{~d}S + \sum \mathrm{~exterior ~angles} & = \left[ \frac{A_\Delta}{R^2} \right] + \bigg[ (\pi-\theta_a) + (\pi-\theta_b) + (\pi-\theta_c) \bigg] \\
    & = \Big[\theta_a + \theta_b + \theta_c - \pi \Big] + \Big[ 3\pi-\theta_a - \theta_b - \theta_c \Big] \\
    & = (\theta_a + \theta_b + \theta_c) - \pi + 3\pi - (\theta_a + \theta_b + \theta_c) \\
    & = 2\pi
\end{align}
thus correlating its differential geometry with its topology. The physical significance of the Gaussian curvature is the intrinsic measure of the surface curvature at a particular point contingent to the surface.

Expanding this notion to a spacetime fabric \cite{3}, the metric of a static, axis-symmetric, asymptotically flat spacetime can be assumed as:
\begin{eqnarray}~\label{metric}
\mathrm{~d} s^{2}=g_{\mu \nu} \mathrm{~d} x^{\mu} \mathrm{~d} x^{\nu}=-f(r) \mathrm{~d} t^{2}+\frac{1}{g(r)} \mathrm{~d} r^{2}+r^{2}\left(\mathrm{~d} \theta^{2}+\sin ^{2} \theta \mathrm{~d} \phi^{2}\right).
\end{eqnarray}

The null geodesics satisfies $\mathrm{~d}s^2=0$, and taking the equatorial plane $
\theta=\pi / 2, \mathrm{~d} \theta=0$, optical metric can be written as:

\begin{equation}
\mathrm{~d} t^{2} \equiv \bar{g}_{i j} \mathrm{~d}x^{i} \mathrm{~d}x^{j}=\bar{g}_{r r} \mathrm{~d}r^{2}+\bar{g}_{\phi \phi} \mathrm{~d}\phi^{2}=\frac{1}{f(r) g(r)} \mathrm{~d}r^{2}+\frac{r^{2}}{f(r)} \mathrm{~d}\phi^{2}
\end{equation}
where, $i$ and $j$ run from $1$ to $3$, and $\gamma_{ij}$.
The Gaussian curvature of the optical metric is
\begin{eqnarray}
\mathcal{K}=\frac{R_{icciScalar}}{2}=-\frac{1}{\sqrt{\bar{g}_{r r} \bar{g}_{\phi \phi}}}\left[\frac{\partial}{\partial r}\left(\frac{1}{\sqrt{\bar{g}_{r r}}} \frac{\partial \sqrt{\bar{g}_{\phi \phi}}}{\partial r}\right)+\frac{\partial}{\partial \phi}\left(\frac{1}{\sqrt{\bar{g}_{\phi \phi}}} \frac{\partial \sqrt{\bar{g}_{r r}}}{\partial \phi}\right)\right].~\label{GC}
\end{eqnarray}

Thus, the Gauss-Bonnet theorem is written as:
\begin{equation}
    \iint_{\mathcal{M}} \mathcal{K} \mathrm{~d} S+\int_{\partial \mathcal{M}} \kappa \mathrm{~d} t+\sum_{i} \alpha_{i}=2 \pi \chi(\mathcal{M})
    \label{gbt}
\end{equation}
where, $\mathcal{M}$ is the selected manifold, $dS$ is a surface element, $\alpha_i$ is the exterior angle at the $i^{th}$ vertex, and $\chi$ is the Euler characteristic of the topology.

*fig*

The extent of deviation of a curve from the shortest length of an arc connecting any two points on a surface is measured by the Geodesic curvature, $\kappa$:

\begin{equation}
\kappa=\frac{1}{2 \sqrt{\bar{g}_{r r} \bar{g}_{\phi \phi}}}\left(\frac{\partial \bar{g}_{\phi \phi}}{\partial r} \frac{\mathrm{~d}\phi}{\mathrm{~d}t}-\frac{\partial \bar{g}_{r r}}{\partial \phi} \frac{\mathrm{~d}r}{\mathrm{~d}t}\right).
\end{equation}

Using bounded $\mathcal{M}$ by a geodesic $C_1$ from the source $S$ to the observer $O$ and a circular curve $C_R$ intersecting $C_1$ in $S$ and $O$ at right angles, Eq.~(\ref{gbt}) reduces to
\begin{eqnarray}
\iint_{\mathcal{M}}\mathcal{K}\mathrm{~d}S+\int_{C_1}\kappa(C_R)\mathrm{~d}t=\pi,~\label{GB2}
\end{eqnarray}
where we have used $\kappa(C_1)=0$ and $\chi(\mathcal{M})=1$. If the center of the lens is singular, then $\chi(\mathcal{M})$ will be taken as zero.

\section{Calculating Deflection Angle using Gibbons and Werner Method}
\label{sgwm}
In order to mitigate the complexity of determining the deflection angle, consider a space domain distinguished by the optical metric for an optical geometry in an asymptotic region with a point source and an observer. Assume that the source, located at infinity from a spherically symmetric distribution of mass or – in this case – the lens, is viewed by the observer located at infinity from the lens (perhaps, a Schwarzschild black hole), in an asymptotically flat spacetime \cite{1}. Analyzing the second integral in Eq.~(\ref{GB2}), the circular curve $C_R := r(\phi)=R=\text{const}$. In the limit $R\rightarrow \infty$, one can obtain,
\begin{align}
    \kappa(C_R) \mathrm{~d}t &= \lim_{R\rightarrow \infty}\left[\kappa(C_R) \mathrm{~d}t\right] \nonumber\\
    &= \lim_{R\rightarrow \infty}\left[\frac{1}{2 \sqrt{\bar{g}_{r r} \bar{g}_{\phi \phi}}}\left(\frac{\partial \bar{g}_{\phi \phi}}{\partial r} \right)\right]\mathrm{~d}\phi \nonumber\\
    &= \mathrm{~d}\phi.~\label{geoR}
\end{align}

Inserting Eq.~(\ref{geoR}) into Eq.~(\ref{GB2}), one has
\begin{eqnarray}
\iint_{\mathcal{M}_{R\rightarrow \infty}}\mathcal{K}\mathrm{~d}S + \int_0^{\pi+\hat{\alpha}}\mathrm{~d}\phi=\pi,
\end{eqnarray}
then, the weak deflection angle can be calculated by integrating its curvature over an infinite region bound by the ray of light apart from the lens:
\begin{eqnarray}
\hat{\alpha}&=&-\iint_{\mathcal{M}}\mathcal{K}\mathrm{~d}S=-\int^{\pi}_0\int^{\infty}_{{u}/{\sin\phi}}\mathcal{K}\mathrm{~d}S~\label{deflang}
\end{eqnarray}
where we have used the zero-order particle trajectory, $r=u/\sin\phi$ and $0\leq\phi\leq\pi$. The distance between the lines passing through the particle (along the direction of its motion) and the center of the gravitating object is known as the impact parameter, $u$. Also,
\begin{equation}
\mathrm{~d}S=\sqrt{\left(\frac{r^2}{f(r)^2 g(r)}\right)} \mathrm{~d} r \mathrm{~d} \phi.
\end{equation}

 This was proposed by Gibbons and Werner \cite{3} as an alternate method to find the deflection angle. 
\subsection{Examples: Black holes}
A black hole is a point of extreme gravity in the spacetime fabric, so strong and 'funneled down' that no particle can escape its pull. When a black hole is formed, there are two possibilities: one, it creates a core that is singular which has a physical solution that emphasizes on the existence of singularity. Here, the Physics we know ceases to thrive due to its infinite density. Beginning with a Schwarzschild black hole, in Eq.~(\ref{metric}):
\begin{equation}
f(r) := 1 - \frac{2M}{r}.
\end{equation}
Applying null geodesics and choosing the equatorial plane $\partial g_{\mu\nu}/\partial \theta = 0$,  the optical metric becomes: 
\begin{equation}\mathrm{~d}t^2 = \frac{\mathrm{~d}r^2}{f(r)^2} + \frac{r^2}{f(r)} \mathrm{~d}\phi^2 \end{equation}
with its Gaussian curvature derived to be:
\begin{equation}
\mathcal{K} = -\frac{2M}{r^3}.  
\end{equation}

Plugging this into Eq.~(\ref{deflang}) with $\mathrm{~d}S \equiv r\mathrm{~d}r\mathrm{~d}\phi$, the zeroth order approximation along a straight line for a Schwarzschild black hole renders the deflected angle \cite{3} to be:
\begin{equation}
    \hat{\alpha} = \frac{4M}{u}.
    \label{sd}
\end{equation}

For the sake of comparison and further understanding, table (\ref{defbht}) is tabulated. It shows how the deflection angle and hence, Eq.~(\ref{sd}) varies in different paradigms.

\begin{center}
\begin{tabular}{ | m{9em} | m{6cm} | m{6cm} | }
  \hline
  \textbf{Case considered} & \textbf{Gaussian curvature} & \textbf{Deflection angle} \\ 
  \hline
  Weyl correction of a Schwarzschild black hole \cite{6} & $$\mathcal{K} = -\frac{2 M}{r^{3}}+\frac{3 M^{2}}{r^{4}}-\frac{72 M \alpha}{r^{5}} $$ & $$\hat{\alpha} = \frac{4 M}{u}+\frac{15 \pi M^{2}}{4 u^{2}}+\frac{32 M \alpha}{u^{3}}+\frac{261 \pi M \alpha}{4 u^{4}}$$ \\ 
  \hline
  Schwarzschild-like solution in Bumblebee gravity \cite{6} & $$\mathcal{K} = -\frac{2 M}{(1+l) r^{3}} $$  & $$\hat{\alpha} = \frac{4 M}{u}+\frac{\pi l}{2}-\frac{2 M l}{u} $$ \\ 
  \hline
  Rindler-modified Schwarzschild black hole \cite{7} & $$\mathcal{K} = -\frac{2 \mu}{r^{3}}\left[f+\frac{\mu}{2 r}-\frac{4 \pi r^{3}}{\mu} F(\rho, f, p, \mu) \right]$$ & $$ \hat{\alpha} = \frac{0.126127529}{\sqrt{a^{3} u^{7}}} $$ \\
  \hline
  Reissner-Nordstr\"om black hole \cite{8} with topological defects & $$\mathcal{K} = -\frac{2 M}{r^{3}}\left(1-\frac{3 M}{2 r}\right)+\mathcal{O}(Q^2, r^4)$$ & $$\hat{\alpha} = \frac{4 M}{u} + 4 \mu \pi  - \frac{3 \pi Q^{2}}{4 u^{2}} + 4\pi^2 \eta^2$$ \\ 
  \hline
  Einstein-Maxwell-Dilaton-Axion \cite{9} (EMDA) black hole & $$\mathcal{K} = - \frac{2M}{r^{3}}+ \frac{3M^{2}}{r^{4}}-\left(\frac{6M}{r^{4}}- \frac{12M^{2}}{r^{5}}\right) r_{0} $$ & $$\hat{\alpha} = \frac{4 {M}}{u}+\frac{3 {M} \pi}{2 u^{2}} $$ \\ 
  \hline
  \cite{10} Extended Uncertainty Principle (EUP) black hole & $$\mathcal{K} = - {\frac {8{M}^{3}\alpha}{{r}^{3}{L}^{2}}} - {\frac {2M}{{r}^{3}}} $$ & $$\hat{\alpha} = \frac{4 M}{u}+\frac{16 \alpha M^{3}}{u L^{2}}$$ \\ 
  \hline
  Regular black holes with cosmic strings (RBCS) \cite{11} & $$\mathcal{K} = -\frac{2 M_{0}}{r^{3}}+\frac{3 M_{0}^{2}}{r^{4}} $$ & $$\hat{\alpha} = \frac{4 M}{u}+4 \pi \mu $$ \\ 
  \hline
  Non linear electrodynamic (NLED) black hole \cite{12} & $$\mathcal{K} = -\frac{2 M}{r^{3}}+\frac{3 Q^{2}}{r^{4}}-\frac{4 M Q \alpha}{r^{3}} $$ & $$\hat{\alpha} = \frac{4 M}{u}-\frac{3 \pi Q^{2}}{4 u^{2}}+\frac{20 M Q \alpha}{u}$$ \\
  \hline
  \end{tabular}
\captionof{table}{Deflection angle caused by various black holes\label{defbht}}
\end{center}

It is worth noticing that anything added to or subtracted from the term $4M/u$ models the deflection angle to the desired case suggesting that the GB method is vastly flexible.

\subsection{Examples: Wormholes}
On the other hand, another possibility during the formation of a black hole is that it forms a bridge through the spacetime fabric to another point (in the same universe or another). In its purest form, the mathematical solution indicates that the center of the compressed matter contains a "throat" called the Einstein-Rosen bridge, technically known as a traversable wormhole.

Table (\ref{defwht}) is analogous to table (\ref{defbht}) but for different types of wormholes:
\begin{center}
\begin{tabular}{ | m{9em} | m{7cm} | m{5cm} | }
  \hline
  \textbf{Case considered} & \textbf{Gaussian curvature} & \textbf{Deflection angle} \\ 
  \hline
  Einstein-Rosen-type wormhole in Weyl gravity \cite{5} & $$\mathcal{K} = \frac{1}{4 U^{4}}- \frac{4M\left(U^{4}+27 \alpha\right)}{U^{10}}+ \mathcal{O}\left(U^{16}\right) $$ & $$\hat{\alpha} = \frac{\pi}{16u^2} + \frac{3\pi M}{8u^4}$$ \\ 
  \hline
  \cite{5} Einstein-Rosen type wormhole in Bumblebee gravity & $$\mathcal{K} = \frac{1}{4 U^{4}(1+l)}- \frac{4M}{U^{6}(1+l)}+ \frac{25M^{2}}{U^{8}(1+l)} $$ & $$\hat{\alpha} = \frac{\pi}{16u^2} + \frac{3\pi M}{8u^4} + \frac{\pi l}{2}$$ \\ 
  \hline
  Einstein-Maxwell-dilaton (charged) wormhole \cite{13} & $$\mathcal{K} = -\frac{16PQ}{r^4} +\frac{\Sigma^2}{r^4}+ \mathcal{O} \left(r^6 \right)$$ & $$\hat{\alpha} = \frac{3 \pi P Q}{2 u^{2}}-\frac{\pi \Sigma^{2}}{4 u^{2}} $$ \\ 
  \hline
  \cite{14} Brane-Dicke wormhole class II \newline \newline \newline Brane-Dicke wormhole class I & $$ \mathcal{K} = - \frac{2\tilde{C} B}{r^{3}} $$ \newline $$ \mathcal{K} = \frac{-e^{\left(\frac{2}{\pi} \arctan \left[\frac{\beta}{r}\right]\right)^{2}}}{\pi^{3} \beta_{0}^{2}}\left[\frac{2 B \pi-4 B \tilde{C} \pi^{2}}{r^{3}}\right] $$ & $$ \hat{\alpha}_{\textrm{II}} = \frac{8 B}{u\sqrt{2\omega + 4}} $$ \newline $$\hat{\alpha}_{\textrm{I}} = \frac{4B}{u}(1-2\pi) + \frac{2\pi B \omega}{u}-\frac{3\pi B \omega^{2} }{4 u} $$ \\
  \hline
  *\cite{15} Wormhole with electric charge \newline \newline \newline Wormhole with a scalar field & $$ \mathcal{K} = \frac{3 Q^{2}-u_{0}^{2}}{r^{4}}-\frac{4 u_{0}^{2} Q^{2}}{r^{6}} $$ \newline $$ \mathcal{K} = \frac{-u_{0}^{2}+\alpha}{r^{4}} $$ & $$\hat{\alpha}_{\textrm{EC}} = \frac{\pi u_{0}^{2}}{4 u^{2}}-\frac{3\pi Q^2}{4 u^{2}} $$ \newline $$\hat{\alpha}_{\textrm{SF}} = \frac{\pi u_{0}^{2}}{4 u^{2}}-\frac{\pi \alpha}{4 u^{2}} $$ \\ 
  \hline
  \cite{16} Black-bounce traversable wormhole & $$ \mathcal{K} = \frac{3M^{2}}{r^{4}}- \frac{2M}{r^{3}}-\frac{a^{2}}{r^{4}} +\mathcal{O} \left(r^5\right) $$ & $$\hat{\alpha} = \frac{4M}{u}+\frac{\pi a^{2}}{4 u^{2}} $$ \\ 
  \hline
  Schwarzschild-like wormhole \cite{17} & $$ \mathcal{K} = -\frac{\left(\lambda^{2}+2\right) M}{\rho^{3} n_{0}^{2}}$$ & $$\hat{\alpha} = \frac{4M}{n_{0} u} + \frac{2M \lambda^{2}}{n_{0} u} $$ \\
  \hline
  Phantom wormhole \cite{18} & $$ \mathcal{K} = \frac{4 r(a-1) r_{0}+(7-14 a) r_{0}^{2}}{4 r^{4}} $$ & $$\hat{\alpha} = \frac{2r_0}{u} (1-a) - \frac{\pi r_0^2}{16u^2}(7-14a) $$ \\ 
  \hline
\end{tabular}
\captionof{table}{Deflection angle caused by various wormholes\label{defwht}}
\end{center}
Like in the case of black holes, a pattern is evident. In most of the cases, the deflection angle is dependent on the term proportional to $\pi/u^2$. In other cases, the inverse dependence of the deflection angle on the impact parameter like in the case of a black hole is observed.

\section{Calculating the Deflection Angle of Non-Asymptotically Flat spacetimes using Finite-Distance Corrections}
\label{sfdm}
One of the underlying assumptions in the derivation of the weak deflection angle is that the distance of both the source and the observer is infinite from the lensing object. On the other hand, every observable object is at a finite redshift from the observer, us. Owing to this, the above equations need a tweak called finite-distance corrections: it is written as the difference between the deflection angles for the asymptotic case and the finite-distance case: \cite{4}
$\delta\hat{\alpha} = \hat{\alpha} - \hat{\alpha}_\infty.$

Ishihara et. al have attained a general form for finite-distance corrections in \cite{19}. Let $\Psi$ be the angle of the light ray measured radially and $\Phi$ be the longitude at a given point \textendash{~subscripts $S$ and $R$ denote the positions of the source and the receiver respectively}. If the coordinate-separation angle between the source and the observer is
$\Phi_{RS} = \Phi_R - \Phi_S$,
then, Eq.~(\ref{deflang}) is formulated as:
\begin{equation}
    \label{fdda}
    \hat{\alpha} = -\iint \mathcal{K} \mathrm{~d}S \equiv \Psi_R - \Psi_S + \Phi_{RS}.
\end{equation}
Correcting this to a finite distance $r$ for both the source and the receiver with $\tilde{r} := 1/r$, then:
\begin{equation}
    \hat{\alpha} = \Psi_R - \Psi_S + \left[ \int_{\tilde{r}_R}^{\tilde{r}_L} \frac{\mathrm{~d} \tilde{r}}{\sqrt{F(\tilde{r})}} + \int_{\tilde{r}_R}^{\tilde{r}_L} \frac{\mathrm{~d} \tilde{r}}{\sqrt{F(\tilde{r})}} \right]
    \label{fde}
\end{equation}
where, the subscript $L$ represents the lens taken to be at zero and,
\begin{equation}
F(\tilde{r}) \equiv \left(\frac{\mathrm{~d}\tilde{r}}{\mathrm{~d}\phi} \right)^2     
\end{equation}
is the photon orbit equation of the spacetime in question. Eq.~(\ref{fde}) is, therefore, corrected for finite-distance.

One application of this method is the Kottler case \cite{19} whose spacetime is given by:
\begin{equation}d s^{2}=-\left(1-\frac{r_{g}}{r}-\frac{\Lambda r^2}{3} \right) d t^{2}+\frac{d r^{2}}{1-\frac{r_{g}}{r}-\frac{\Lambda r^2}{3}}+r^{2}\left(d \theta^{2}+\sin ^{2} \theta d \phi^{2}\right)\end{equation}
where $r_g$ is a geodesic parameter in the metric and $\Lambda$ is the cosmological constant. The orbit equation for the Kottler spacetime, also known as the Schwarzschild de-Sitter spacetime, is defined as:
\begin{equation}F(\tilde{r})=\frac{1}{u^{2}}-\tilde{r}^{2}+r_{g} \tilde{r}^{3}+\frac{\Lambda}{3}.\end{equation}
Expanding over $r_g$ and $\Lambda$ to find $\Psi_R$ and $\Psi_S$ followed by determining $\Phi_{RS}$, the deflection angle for the non-asymptotically flat case is found to be:
\begin{equation}\hat{\alpha} = \frac{r_{g}}{u}\left[\sqrt{1-u^{2} \tilde{r}_{R}^{2}}+\sqrt{1-u^{2} \tilde{r}_{S}^{2}}\right]-\frac{\Lambda u}{6}\left[\frac{\sqrt{1-u^{2} \tilde{r}_{R}^{2}}}{\tilde{r}_{R}}+\frac{\sqrt{1-u^{2} \tilde{r}_{S}^{2}}}{\tilde{r}_{S}}\right]+\frac{r_{g} \Lambda u}{12}\left[\frac{1}{\sqrt{1-u^{2} \tilde{r}_{R}^{2}}}+\frac{1}{\sqrt{1-u^{2} \tilde{r}_{S}^{2}}}\right].\end{equation}

Encompassing the scope of the source and the observer to positions far from the lens, the deflection angle reduces to:
\begin{equation}\hat{\alpha} = \frac{2 r_{g}}{u}-\frac{\Lambda u}{6} \left(\frac{1}{\tilde{r}_{R}}+\frac{1}{\tilde{r}_{S}}\right)+\frac{r_g \Lambda u}{6} \end{equation}
as derived by \cite{19} which agrees with \cite{3}.

For a non-asymptotically flat spacetime, it is intriguing to notice that under the case of Weyl conformal gravity, the deflection angle is derived \cite{19} to be:
\begin{equation}
\hat{\alpha} = \frac{2M}{u}\left(\sqrt{1-u^{2} \tilde{r}_{R}^{2}}+\sqrt{1-u^{2} \tilde{r}_{S}^{2}}\right)
- \gamma M \left(\frac{u \,\tilde{r}_{R}}{\sqrt{1-u^{2} \tilde{r}_{R}^{2}}}+ \frac{u \,\tilde{r}_{S}}{\sqrt{1-u^{2} \tilde{r}_{S}^{2}}}\right)
\label{dafdc}
\end{equation}
where, $\gamma$ is a metric parameter. Extrapolating this equation of the deflection angle at a finite-distance, when the source and the observer are too far from the lens such that $u\,\tilde{r}_R \ll 1$ and $u\,\tilde{r}_R \ll 1$, it remarkably reduces to Eq.~(\ref{sd}) with negligible higher-order terms.

\subsection{Examples: Black holes}
Although the above-mentioned method has demonstrated reliability, it is not suitable for black hole due to the singularity of the lens. Ono and Asada \cite{4} have extended this to strong deflection limits to include loops in the photon orbits and for Sagittarius A*. The latter is evaluated to be $\sim 10^{-5} \,\textrm{arcsec}$, much larger than the corrections required for M87 since Sagittarius A* is comparatively closer to us.

In \cite{2}, the authors have studied the case of a Kerr black hole with Modified Gravity alterations. They have corrected for finite-distance pertaining to the fact that the deflection angle depends not only on the Gaussian curvature but also the geodesic curvature. So, the deflection angle with finite-distance corrections is expressed as:
\begin{equation}
    \hat{\alpha} = -\iint \mathcal{K} \mathrm{~d}S - \int \kappa \mathrm{~d}\ell
    \label{dp}
\end{equation}
where, $\ell$ is the arc length of the photon sphere and the line element $\mathrm{~d}\ell^2 \equiv \mathrm{~d}r^2 + r^2 \left(\mathrm{~d}\theta^2 + \sin^2{\theta}\mathrm{~d}\phi^2\right) $ gives rise to the path integral \cite{2} in the second term of Eq.~(\ref{dp}).
\newline The resulting deflection angle has been found to be:
\begin{equation}
    \hat{\alpha} = \frac{4M}{u}(1+\alpha)\pm \frac{4aM}{u^2} (1+\alpha)
    \label{kerrfd}
\end{equation}
where, $\alpha$ is a parameter that determines the degree of deviation of MoG from General Relativity in addition to governing the strength of gravity and $a$ is the spin parameter that corresponds to a Kerr black hole. The result of this is non-trivial as shown by their plots.

\subsection{Examples: Wormholes}
Consider the case of a Teo wormhole, an axially symmetric rotating wormhole which has the most generic solution for a traversable wormhole. The deflection angle is given by:
\begin{equation}
    \hat{\alpha} = \frac{u_{0}}{u} \pm \frac{4 a}{u^{2}}.
    \label{teows}
\end{equation}
But when corrected for finite-distance, the deflection angle becomes:
\begin{equation}\hat{\alpha} = \frac{u_{0}}{2 u}\left(\sqrt{1-u^{2} \tilde{r}_{R}^{2}} + \sqrt{1-u^{2} \tilde{r}_{S}^{2}}\right) \pm \frac{2 a}{u^{2}} \left(\sqrt{1-u^{2} \tilde{r}_{S}^{2}}+\sqrt{1-u^{2} \tilde{r}_{R}^{2}}\right) \end{equation}
which is in agreement with Eq.~(\ref{dafdc}), with the positive sign signifying retrograde and the negative sign signifying prograde.\cite{4}

\section{Calculating Deflection Angle in Plasma Medium}
\label{spm}
So far, the discussion was concentrated on the light rays travelling through vacuum medium. Next step is to probe into a medium of plasma; observations of the Event Horizon Telescope \cite{20} necessitate this now more than ever. When matter falls into a black hole, it is essentially disintegrated by the extreme gravitational force of the black hole which then heated up to millions of degrees creating magnetized plasma. 

When a light ray encounters this hot soup of ionized gas, it will experience refraction, implying that there is a factor of refractive index involved, written as:
\begin{equation}n(r) = \sqrt{1 - \frac{\omega_e^2}{\omega_\infty^2} \left(1-\frac{2M}{r}\right)}\end{equation}
where,  $\omega_e$ is the plasma frequency of an electron and $\omega_\infty$ is the photon frequency both measured at infinity \cite{21}. This non-zero factor causes the rays of light that pass by to be significantly deflected. To find the deflection angle on the plasma medium, one should use the corresponding optical metric is:

\begin{equation}
d \sigma ^ { 2 } = g _ { i j } ^ { \mathrm { opt } } d x ^ { i } d x ^ { j } = \frac { n ^ { 2 } ( r ) } { f ( r ) } \left( \frac{d r ^ { 2 }}{f(r)} +  r^2 d \varphi ^ { 2 }\right).
\end{equation}

\subsection{Examples: Black holes}
The deflection angle of a Schwarzschild black hole in a plasma medium is given by \cite{21}:
\begin{equation}\hat{\alpha}=\frac{2M}{u} \left[ 1 + \frac{1}{1 - \left(\omega_e^2/\omega_\infty^2\right)}\right].\end{equation}

Going back to the second half of table (\ref{defbht}), let us examine the effect of plasma on the corresponding black holes and re-tabulate in table (\ref{defbpt}).

The contribution of the plasma term is apparent in every case such that if $\omega_e/\omega_\infty \rightarrow 0$, all of the above expressions boil down to their non-plasmic counterparts. Conclusively, the influence of plasma is distinctly discernible, as shown graphically by \cite{22}.

\begin{center}
\begin{tabular}{ | m{5em} | m{7.1cm} | m{7cm} | }
  \hline
  \textbf{Case} & \textbf{Gaussian curvature} & \textbf{Deflection angle} \\ 
  \hline
  EMDA black hole \cite{9} & $$\mathcal{K} = \frac{M\left(\omega_{e}^{2} r-2 \omega_{\infty}^{2} r+4 r_{0} \omega_{e}^{2}-6 r_{0} \omega_{\infty}^{2}\right) \omega_{\infty}^{2}}{\left(\omega_{e}^{2}-\omega_{\infty}^{2}\right)^{2} r^{4}} $$ & $$\hat{\alpha} = \frac{4M}{u} + \frac{3M r_{0} \pi}{2u^{2}} + \frac{4M}{u} \frac{\omega_{e}^{2}}{\omega_{\infty}^{2}} + \frac{5\pi M r_{0}}{4u^{2}} \frac{\omega_{e}^{2}}{\omega_{\infty}^{2}} $$ \\ 
  \hline
  EUP black hole \cite{10} & $$\mathcal{K} = -{\frac {2M}{{r}^{3}}}+{\frac {3{M}^{2}}{{r}^{4}}}-{\frac  {8{M}^{3}\alpha}{{L}^{2}{r}^{3}}} -{\frac {M\omega_e^2}{{\omega_\infty^2}r^3}} $$ $$ + {\frac {4{M}^{2}\omega_e^2}{{\omega_\infty^2}{r}^{4}}} -\left( {\frac {4\alpha}{{\omega_\infty^2}{L}^{2}{r}^{3}}}+ {\frac {4}{{\omega_\infty^2}{r}^{5}}} \right) {M}^{3}  {\omega_e^2} $$ & $$\hat{\alpha} = \frac{4M}{u}+\frac{16 \alpha M^{3}}{u L^{2}}+\frac{6M}{u}\frac{\omega_{e}^{2}}{\omega_{\infty}^{2}}+\frac{24 \alpha L^{2} M^{3}}{u} \frac{\omega_{e}^{2}}{\omega_{\infty}^{2}} $$  \\ 
  \hline
  RBCS \cite{11} & $$\mathcal{K} = \frac{M\left(\omega_{e}^{2}-2 \omega_{\infty}^{2}\right) \omega_{\infty}^{2}}{\left(\omega_{e}^{2}-\omega_{\infty}^{2}\right)^{2} r^{3}}- \frac{3M^{2}\left(\omega_{e}^{2}+\omega_{\infty}^{2}\right) \omega_{\infty}^{4}}{\left(\omega_{e}^{2}-\omega_{\infty}^{2}\right)^{3} r^{4}} $$ & $$\hat{\alpha} = \frac{4 M}{u} + 4 \pi \mu + \frac{4M}{u}\frac{\omega_{e}^{2}}{\omega_{\infty}^{2}} $$ \\ 
  \hline
  NLED black hole \cite{12} & $$\mathcal{K} = \frac{M}{r^{3}}\left(-2-\frac{\omega_{e}^{2}}{\omega_{\infty}^{2}}+\frac{2 \omega_{e}^{4}}{\omega_{\infty}^{4}}\right)+\frac{2 M Q^{2}}{r^{5}}$$ $$\left(1-\frac{17 \omega_{e}^{2}}{\omega_{\infty}^{2}}+\frac{5 \omega_{e}^{4}}{\omega_{\infty}^{4}}\right) -\frac{4 M Q \alpha}{r^{3}}\left(1+\frac{\omega_{e}^{2}}{\omega_{\infty}^{2}}- \frac{3\omega_{e}^{4}}{\omega_{\infty}^{4}}\right) $$ & $$\hat{\alpha} = \frac{4M}{u} - \frac{3 Q^{2} \pi}{4 u^{2}} + \frac{4MQ \alpha}{u} + \frac{2M}{u} \frac{\omega_{e}^{2}}{\omega_{\infty}^{2}} $$ \quad \quad $$ + \frac{2MQ \alpha}{u}\frac{\omega_{e}^{2}}{\omega_{\infty}^{2}} -\frac{6M}{u}\frac{\omega_{e}^{4}}{\omega_{\infty}^{4}} - \frac{3 Q^{2} \pi}{4u^{2}} \frac{\omega_{e}^{4}}{\omega_{\infty}^{4}} $$ \\ 
  \hline
\end{tabular}
\captionof{table}{Deflection angle caused by various black holes in the presence of homogenous plasma\label{defbpt}}
\end{center}

\subsection{Examples: Wormholes}
Similarly, plasma medium contributes conspicuously to the deflection angle due to a wormhole. As discussed in \cite{23}, the Gaussian curvature and the deflection angle of a Casimir wormhole \textendash ~a type of a travesable wormhole generated by the Casimir effect \textendash ~is found to be:
\begin{equation} \mathcal{K} =  -\frac{2 a}{3 r^{3}}+\frac{2 a^{2}}{3 r^{4}} \quad \quad \textrm{(and)} \end{equation}
\begin{equation}\hat{\alpha} = \frac{4 a}{3 u}-\frac{\pi a^{2}}{6 u^{2}}\end{equation}
respectively. Casimir effect is a physical attractive force between two parallel, conducting boundaries caused by quantum field fluctuations which allows the energy to be relatively negative at a specific point. Incorporating the effects of plasma, the Gaussian curvature becomes:
\begin{equation} \mathcal{K} = -\frac{\omega_{e}^{2} a}{\omega_{\infty}^{2} r^{3}}-\frac{2}{3} \frac{a}{r^{3}}+ \frac{7}{3} \frac{a^{2} \omega_{e}^{2}}{r^{4} \omega_{\infty}^{2}}+ \frac{2}{3} \frac{a^{2}}{r^{4}} \end{equation}
to give the deflection angle:
\begin{equation}\hat{\alpha} = \frac{4 a}{3 u}-\frac{\pi a^{2}}{6 u^{2}} + \frac{2 a}{u} \frac{\omega_{e}^{2}}{\omega_{\infty}^{2}} - \frac{7 \pi a^{2}}{12 u^{2}} \frac{\omega_{e}^{2}}{\omega_{\infty}^{2}}\end{equation}
hence sustaining its property of distinctive flexibility.

Examining these analyses, it is evident from the photons' equation of motion for a wormhole in homogenous plasma that the inner radius of the photon orbits around the wormhole decreases with the existence of plasma.

\section{Calculating Deflection Angle of Stationary Black holes}
\label{ssbh}
A non-rotating black hole with no charge or time-dependence and exhibiting axis-symmetry is categorized as a stationary black hole. The periphery of such a black hole from which nothing can escape, known as the event horizon, is non-expanding. The Schwarzschild spacetime is independent of time, and hence, stationary.

The line element of one such spacetime is given by \cite{1}:
\begin{equation}\mathrm{~d} s^{2}=-A(r, \theta) \mathrm{~d} t^{2}-2 H(r, \theta) \mathrm{~d} t \mathrm{~d} \phi + B(r, \theta) \mathrm{~d} r^{2}+C(r, \theta) \mathrm{~d} \theta^{2}+D(r, \theta) \mathrm{~d} \phi^{2}.\end{equation}
At the equatorial plane, applying null condition:
\begin{align}
    \mathrm{~d} t &= \sqrt{\gamma_{i j} \mathrm{~d} x^{i} \mathrm{~d} x^{j}}+\beta_{i} \mathrm{~d} x^{i} \\
    &\equiv \frac{B(r, \theta)}{A(r, \theta)} \mathrm{~d} r^{2}+\frac{C(r, \theta)}{A(r, \theta)} \mathrm{~d} \theta^{2}+\frac{A(r, \theta) D(r, \theta)+H^{2}(r, \theta)}{A^{2}(r, \theta)} \mathrm{~d} \phi^{2}-\frac{H(r, \theta)}{A(r, \theta)} \mathrm{~d} \phi
\end{align}
where, $\gamma_{i j}$ is the spatial metric (not $g_{i j}$) defined as the arc length along the photon orbit.

Ergo, the Gauss-Bonnet theorem can be expressed as:
\begin{equation}\iint_{{}_R^{R_{\infty}} \square_{S}^{S_{\infty}}} K \mathrm{~d} S+\int_{R}^{S} \kappa_{g} \mathrm{~d} \ell+\int_{S_{\infty}}^{R_{\infty}} \bar{\kappa}_{g} \mathrm{~d} \ell+\left[\Psi_{R}+\left(\pi-\Psi_{S}\right)+\pi\right]=2 \pi.\end{equation}

The limit of the double integral ${}_R^{R_{\infty}} \square_{S}^{S_{\infty}}$ is a quadrilateral on the equatorial plane embedded on to $\gamma_{i j}$ which is the associated 3-dimensional space. The geodesic curvatures of the paths from $R$ to $R_\infty$ and $S$ to $S_\infty$ are zero owing to the geodesic nature, $\kappa_g$ is the photon geodesic curvature. $\bar{\kappa}_g$ is the geodesic curvature of the segment of circular arc with infinite radius, $R_\infty = S_\infty$.

\subsection{Deflection Angle using the Werner Method for  stationary spacetime}

One of the solutions to a stationary black hole is  the Kerr solution. Werner employed a new geometric approach in \cite{24} to determine the deflection of light. The line element of the Kerr spacetime (with the rising Randers metric) in the Boyer-Lindquist coordinates is written as: \begin{equation}\mathrm{d} s^{2}=-\frac{\Delta}{\rho^{2}}\left(\mathrm{~d} t - a \sin ^{2} \theta \mathrm{~d} \phi\right)^{2}+\frac{\sin ^{2} \theta}{\rho^{2}}\left[\left(r^{2}+a^{2}\right) \mathrm{d} \varphi-a \mathrm{~d} t\right]^{2}+\frac{\rho^{2}}{\Delta} \mathrm{~d} r^{2}+\rho^{2} \mathrm{~d} \theta^{2} \end{equation}
where,
\begin{align}
    \Delta &= r^{2}-2 m r+a^{2} \quad \mathrm{(and)} \\ 
    \rho^{2} &= r^{2}+a^{2} \cos ^{2} \theta.
\end{align}

The metric components of the osculating Riemannian manifold to the first order are:
\begin{align}
    \bar{g}_{r r} &= 1+\frac{4 m}{r}-\frac{2 mar}{b^{3}} \frac{\sin ^{6} \phi}{\left(\cos ^{2} \phi+\frac{r^{2}}{b^{2}} \sin ^{4} \phi\right)^{3/2}} \\
    \bar{g}_{r \phi} &= \frac{2 m a}{r} \frac{\cos ^{3} \phi}{\left(\cos ^{2} \phi+\frac{r^{2}}{b^{2}} \sin ^{4} \phi\right)^{3/2}} \\
    \bar{g}_{\varphi \varphi} &= r^{2}+2 m r-\frac{2 m a r}{b} \frac{\sin ^{2} \phi\left(3 \cos ^{2} \phi+2 \frac{r^{2}}{b^{2}} \sin ^{4} \phi\right)}{\left(\cos ^{2} \phi+\frac{r^{2}}{b^{2}} \sin ^{4} \phi\right)^{3/2}}
\end{align}
with its determinant calculated to be:
\begin{equation} \mathrm{det} ~\bar{g}=r^{2}+6 m r-\frac{6 mar}{b} \frac{\sin ^{2} \phi}{\sqrt{\cos ^{2} \phi+\frac{r^{2}}{b^{2}} \sin ^{4} \phi}}. \end{equation}
Solving for the Christoffel symbols and the Ricci tensor, the Gaussian curvature is obtained to be:
\begin{equation}\mathcal{K} = -\frac{2M}{r^{3}}+\frac{3M a}{u^{2} r^{2}} f(r, \phi) \end{equation}
where,

\begin{eqnarray}
     f(r, \phi) &=\frac{\sin ^{3} \phi}{\left(\cos ^{2} \phi+\frac{r^{2}}{b^{2}} \sin ^{4} \phi\right)^{7/2}}\left[2 \cos ^{6} \phi\left(\frac{5r}{b} \sin \phi-2\right)+\cos ^{4} \phi \sin ^{2} \phi\left(\frac{9r}{b} \sin \phi- \frac{10r^{3}}{b^{3}} \sin ^{3} \phi-2\right)\right. \notag\\
     & \left.+ \frac{2r}{b} \cos ^{2} \phi \sin ^{5} \phi\left(2-\frac{r^{2}}{b^{2}}+\frac{r^{2}}{b^{2}} \cos 2 \phi+ \frac{4r}{b} \sin \phi\right)+\frac{r^{2}}{b^{2}}\left(-\frac{r}{b} \sin ^{9} \phi+ \frac{2r^{3}}{b^{3}} \sin ^{11} \phi+\sin ^{4} 2 \phi\right)\right].
\end{eqnarray}

Hence, the deflection angle in a Kerr-Randers optical geometry is:
\begin{equation}
    \hat{\alpha} = \frac{4M}{u} \pm \frac{4aM}{u^2}
    \label{kerrs}
\end{equation}
which resembles Eq.~(\ref{kerrfd}) derived for a Kerr-MoG black hole; $a$ is the angular momentum parameter.

Jusufi and \"Ovg\"un have applied this method to a rotating global monopole spacetime in \cite{l56}. With the defined line element and the correcsponding expressions for the metric, the Christoffel symbols and the Ricci tensor, the Gaussian curvature is derived to be:
\begin{equation}\mathcal{K}=-\frac{2 M}{r^{3}}+\frac{3 M a}{r^{2}} f(r, \phi, \beta)\end{equation}
where,
 \begin{eqnarray}
     f(r, \phi, \beta) &= \frac{\sin ^{3} \phi}{b^{7}\left(\cos ^{2} \phi+\frac{r^{2} \beta^{2} \sin ^{4} \phi}{b^{2}}\right)^{7/2}}\left(2 \beta^{6} r^{5} \sin ^{11} \phi+5 \beta^{4} b^{2} r^{3} \cos ^{2} \phi \sin ^{7} \phi-10 \beta^{2} b^{2} r^{3} \cos ^{4} \phi \sin ^{5} \phi\right.\notag \\
     &-9 \beta^{2} b^{2} r^{3} \cos ^{2} \varphi \sin ^{7} \phi-\beta^{2} r^{3} b^{2} \sin ^{9} \phi+16 \beta^{2} b^{3} r^{2} \cos ^{4} \phi \sin ^{4} \phi+8 \beta^{2} b^{3} r^{2} \cos ^{2} \phi \sin ^{6} \phi \notag \\
     &-2 \beta^{2} b^{4} r \cos ^{4} \phi \sin ^{3} \phi+10 b^{4} r \cos ^{6} \varphi \sin \phi+11 b^{4} r \cos ^{4} \phi \sin ^{3} \phi+4 b^{4} r \cos ^{2} \phi \sin ^{5} \phi\notag \\
     &\left.-4 b^{5} \cos ^{6} \phi-2 b^{5} \cos ^{4} \phi \sin ^{2} \phi\right).
\end{eqnarray} 
Defining $\eta$ as the scale of gauge-symmetry breaking that is associated with the spacetime of the global monopole, the deflection angle is derived to be:
\begin{equation}\hat{\alpha} = \frac{4 M}{b} \pm \frac{4 M a}{b^{2}}+4 \pi^{2} \eta^{2}+\frac{16 \pi M \eta^{2}}{b}.\end{equation}

They have also analyzed the case of a rotating Letelier spacetime \cite{l56} for which the Gaussian curvature is calculated to be:
\begin{equation} \mathcal{K} = -\frac{2 M}{r^{3}}+\frac{3 M a}{r^{2}} f(r, \phi, A) \end{equation}
where, $f(r, \phi, A)$ is the same as $f(r, \phi, \beta)$ for $\beta = \sqrt{1-A}$ giving:
\begin{equation}\hat{\alpha} = \frac{4 M}{b} \pm \frac{4 M a}{b^{2}}+\frac{\pi A}{2}+\frac{2 M A}{b}.\end{equation}

Analogizing this method given by Werner with \cite{25} in which the authors have plotted Eq.~(\ref{kerrs}) against a rotating Teo wormhole whose deflection angle is found to be Eq.~(\ref{teows}), it is obvious that although the behaviors of the deflection angles of the Teo wormhole and the Kerr solution are very much alike, the light deflection is stronger in the latter case.

\subsection{Deflection Angle using the finite-distance corrections in stationary axisymmetric spacetime}

An intriguing attempt of applying the finite-distance corrections to stationary black holes is presented by \cite{4}. Revisiting the definition of the tweak corresponding to finite-distance corrections:
\begin{equation} \delta \hat{\alpha} = \mathcal{O} \left(\frac{J}{\tilde{r}_S^2}, \frac{J}{\tilde{r}_R^2}\right) \end{equation}
where, $J$ is the spin angular momentum of the lens. This is akin to the second post-Newtonian effect with the factor of the spin parameter. Here, it is prominent that the deflection angle is independent of the impact parameter. Otherwise, the deflection angle would be:
\begin{equation} \hat{\alpha} = \frac{4M}{u} - \frac{4aM}{u^2} + \frac{15\pi M^2}{4u^2} \end{equation}
at the infinite distance limit with the finite-distance corrections for a Kerr black hole. The last term of this equation is the second-order Schwarzschild contribution to the deflection angle, without which the above equation agrees with Eq.~(\ref{kerrs}).

\section{Calculating Deflection Angle using Jacobi Metric Approach within GBT}
\label{sjma}
Contemplating on Eq.~(\ref{dp}), the geodesic curvature of a non-geodesic circular (photon) orbit around the lens causing deflection must be inspected. In order to avoid the geodesic curvature term, the authors of \cite{26} have used a geodesic circular orbit, and then employing the GBT to find the deflection angle.

Jacobi metric is utilized to derive the radius of the circular orbit using a geometric method for a particle in the equatorial plane. For the line element,
\begin{equation}\mathrm{~d}s^2 = -A(r) \mathrm{~d}t^{2} + B(r) \mathrm{~d}r^{2} + C(r) \mathrm{~d}\Omega^{2}, \end{equation}
then its Jacobi metric is written as:
\begin{equation}\mathrm{~d}l^{2}=m^{2}\left(\frac{1}{1-v^{2}}-A\right)\left(\frac{B}{A} \mathrm{~d}r^{2}+\frac{C}{A} \mathrm{~d}\phi^{2}\right).\end{equation}
Taking the particle velocity to be unity, the orbit equation is:
\begin{equation}\left(\frac{\mathrm{~d} \tilde{r}}{\mathrm{~d} \phi}\right)^{2}=\frac{C^{4} \tilde{r}^{4}}{A B} \left(\frac{1}{u^{2}}-\frac{A}{C}\right).\end{equation}

Implementing Eq.~(\ref{fdda}) with the facet that the circular orbit is perpendicular to the outgoing radial lines of light rays, the deflection angle is formulated as:
\begin{equation} \hat{\alpha} = \iint_\mathcal{M} \mathcal{K} \mathrm{~d}S + \Phi_{RS}. \end{equation}

Re-examining the deflection angle in the Weyl and  Bumblebee gravities as in table (\ref{defbht}), the Gaussian curvatures are:
\begin{equation}\mathcal{K}_{W} = -\frac{\gamma^{2}}{4}-\frac{(2+3 r \gamma) M}{r^{3}}+\frac{3(1+2 r \gamma) M^{2}}{r^{4}} \quad \textrm{(and)} \end{equation}\begin{equation} \mathcal{K}_{B} = \frac{M\left(1-v^{2}\right)}{m^{2} r^{3} \lambda^{2}} \frac{\left[8 M^{3}\left(1-v^{2}\right)^{2}-r^{3} v^{2}\left(1+v^{2}\right)-3 M r^{2} v^{2}\left(1-2 v^{2}\right)-6 M^{2} r\left(1-3 v^{2}+2 v^{4}\right)\right]}{\left[r v^{2}+2 M\left(1-v^{2}\right)\right]^{3}} \end{equation}
where, $\gamma$ is a metric constant and $\lambda = \sqrt{1+l}$, for $l$ is the Lorentz violation parameter in a static, spherically symmetric, asymptotically non-flat spacetime with the lens at finite distance. The corresponding deflection angles are:
\begin{equation} \hat{\alpha}_{W} = \frac{2 M\left(\sqrt{1-u^{2} \tilde{r}_{R}^{2}}+\sqrt{1-u^{2} \tilde{r}_{S}^{2}}\right)}{u}-M u \gamma\left(\frac{\tilde{r}_{R}}{\sqrt{1-u^{2} \tilde{r}_{R}^{2}}}+\frac{\tilde{r}_{S}}{\sqrt{1-u^{2} \tilde{r}_{S}^{2}}}\right) \quad \textrm{(and)} \end{equation}
\begin{equation} \hat{\alpha}_{B} = (\lambda-1)\left(\pi-\sin^{-1} u \tilde{r}_{R} - \sin^{-1} u \tilde{r}_{S}\right) + \left[\frac{\left(1+v^{2}\right) \lambda-u^{2} \tilde{r}_{R}^{2}\left(1+v^{2} \lambda\right)}{\sqrt{1-u^{2} \tilde{r}_{R}^{2}}}+\frac{\left(1+v^{2}\right) \lambda-u^{2} \tilde{r}_{S}^{2}\left(1+v^{2} \lambda\right)}{\sqrt{1-u^{2} \tilde{r}_{S}^{2}}}\right] \frac{M}{u v^{2}}. \end{equation}
For the source and observer at infinity, the last equation shrinks to:
\begin{equation} \hat{\alpha} = \pi (\lambda-1) + \frac{2 \lambda M\left(1+v^{2}\right)}{u v^{2}} \end{equation}
which is in agreement with \cite{3} using the conventional formula:
\begin{equation} \hat{\alpha} = \iint_{{{}_R^\infty} \square_{S}^{\infty}} \mathcal{K} \mathrm{~d} S + \int_{S}^{R} \kappa_{g} \mathrm{~d} \sigma+\left(1-\frac{1}{\lambda}\right) \Phi_{RS}. \end{equation}

As for the case of wormholes, \cite{27} can be referred to find the deflection angle of light of a Janis-Newman-Winnicour (JNW) wormhole using the Jacobi metric method:
\begin{equation}\hat{\alpha} = \frac{4\tilde{\gamma} M}{u}+\frac{\left(16 \tilde{\gamma}^{2}-1\right) \pi M^{2}}{4 u^{2}} \end{equation}
where, $\tilde{\gamma}$ is the ratio of the mass related to the asymptotic scalar charge to the mass of the wormhole. The leading term is consistent with the expressions found earlier.

Another illustration discussed in \cite{27} is the charged Einstein-Maxwell-dilaton wormhole. The deflection angle is derived to be:
\begin{equation}\hat{\alpha} = \frac{3 \pi P Q}{2 u^{2}}-\frac{\pi \Sigma^{2}}{4 u^{2}}-\frac{15 \pi P Q \Sigma^{2}}{16 u^{4}}+\frac{105 \pi P^{2} Q^{2}}{16 u^{4}}\end{equation}
which is congruous to the results in table (\ref{defwht}). In both of the above examples, the velocity, $v$, is taken to be 1 since we are heeding to the case of light.

\section{Conclusion}
\label{conc}
In this review, we have brought quite a few researches together which talk about finding deflection angle using the Gauss-Bonnet theorem. The aim was to summarize the expressions of the deflection angle for various cases in an attempt to compare and contrast one another, possibly spotting a pattern. It was noticed that every adjunct had a unique and distinct role in modifying the deflection angle for all the cases considered.

Initially, the Gauss-Bonnet theorem was examined, followed by studying the Gibbons and Werner method to calculate the deflection angle with Eq.~(\ref{deflang}). These were analyzed for a few black holes in table (\ref{defbht}) and wormholes in table (\ref{defwht}). Then, the source and the observer were taken to be at a finite distance and this correction gave rise to an extensive equations which reduced to the established equations at infinite limits; this was done for both black holes and wormholes as well. Next, the effect of inducing a medium of ionized plasma was investigated: yet again, the plasmic terms manifested discretely in both black holes and wormholes such that its contribution reduced the deflection angle to the case of vacuum when removed. Stationary black holes were scrutinized for a Kerr black hole and with finite-distance corrections, only to beget a coalesced expression for the deflection angle. Lastly, the Jacobi metric was perused for black holes and wormholes, which yielded coherent results. 
\\ \\
\textbf{Acknowledgments}
\\ \\
A. \"{O} truly appreciates Professor Alikram Aliev from the Scientific and Technological Research Council of Turkey (TÜBİTAK) and the Turkish Journal of Physics for invitation to write this review article.

\end{document}

%% file: fizik.tex
\newcommand{\bc}{\begin{center}}
\newcommand{\ec}{\end{center}}

\numberwithin{equation}{section}

\renewcommand{\phi}{\varphi}